\documentclass[prl,10pt,notitlepage,twocolumn]{revtex4-1}

\usepackage{graphicx}
\usepackage{amsmath}
\usepackage{amssymb}
\usepackage{amsfonts}
\usepackage{color}
\usepackage{xspace}
\usepackage{subfigure}
\usepackage{bm}
\usepackage{bbold}

\newcommand{\hyperhoney}{[(C$_2$H$_5$)$_3$NH]$_2$Cu$_2$(C$_2$O$_4$)$_3$\xspace}
\newcommand{\hhc}{\hyperhoney}

\begin{document}

\title{A theory of the quantum spin liquid in the hyper-honeycomb metal-organic framework  {\hhc} from first principles}
\author{A. C. Jacko }\affiliation{School of Mathematics and Physics, The University of Queensland, Brisbane, Queensland, 4072, Australia}
\author{B. J. Powell}\affiliation{School of Mathematics and Physics, The University of Queensland, Brisbane, Queensland, 4072, Australia}

\begin{abstract}
We construct a tight-binding model of \hhc from Wannier orbital overlaps. 
Including interactions within the Jahn-Teller distorted Cu-centered $e_g$ Wannier orbitals leads to an effective Heisenberg model. 
The hyper-honeycomb lattice contains two symmetry distinct sublattices of Cu atoms arranged in coupled chains. 
One sublattice is strongly dimerized, the other forms isotropic antiferromagnetic chains. 
Integrating out the strongest (intradimer) exchange interactions leaves extremely weakly coupled Heisenberg chains, consistent with the observed low temperature physics.
\end{abstract}

\maketitle

The idea that some magnetic materials may have quantum disordered ground states at absolute zero \cite{AndersonRVB} has driven extensive efforts to find and understand these quantum spin liquids \cite{Savary}.  Mechanisms that may lead to disordered ground states include frustrated spin interactions (e.g., on kagome \cite{kagome} and anisotropic triangular \cite{RRP} lattices), highly anisotropic spin interactions (e.g., Kitaev interactions on the honeycomb \cite{Kitaev,Jackeli} and hyper-honeycomb \cite{hyperhoneycomb} lattices), ring exchange \cite{Montrunich,MerinoHolt}, and quasi-one-dimensionality \cite{Schulz,Boquet,Kohno}. Nevertheless, relatively few experimental realizations of quantum spin liquids have been found \cite{Balents,Savary,RRP}. Nor is there yet direct experimental evidence for the long-range entanglement that is thought to characterize many quantum spin liquids \cite{Savary}.

\begin{figure}
	\begin{center}
		\includegraphics[width=0.9\columnwidth]{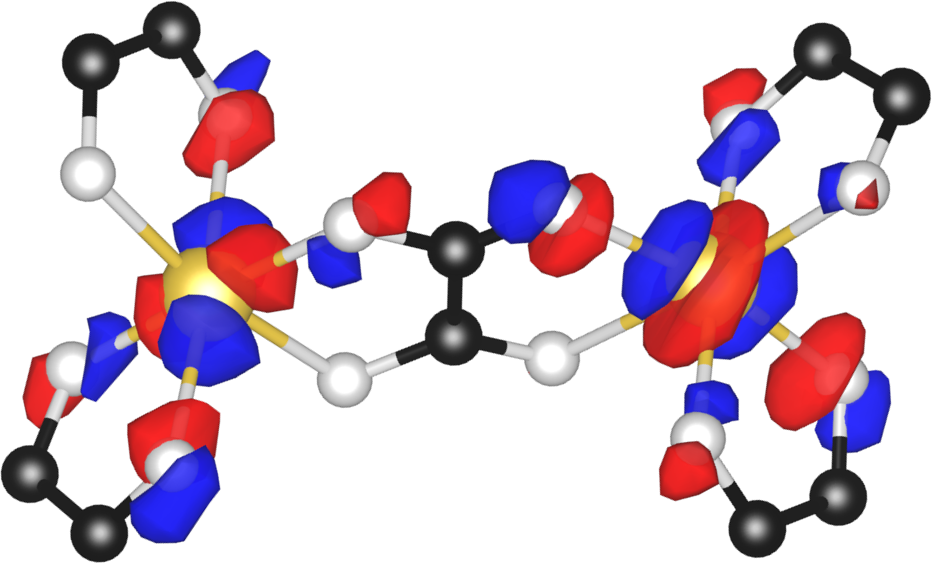} 
		\caption{Two of the Wannier functions of \hhc, shown on adjacent copper sites; both extend out anisotropically onto the oxalate bridging ligands. The left-hand orbital is a d$_{x^2-y^2}$-type orbital, while the right-hand orbital is largely d$_{z^2}$, the $e_g$ partner of d$_{x^2-y^2}$. 
		White (black) atoms are oxygen (carbon). }\label{fig:WF}
	\end{center}
\end{figure}

Given the number of theoretical proposals for quantum disordered magnetic ground states, it is interesting to ask whether one might design new materials to realize specific ideas from the bottom up. Clearly, this is a highly non-trivial objective. But, the goal of rationally designing new materials is now actively considered in several of areas of chemistry. Further encouragement comes from recent progress in quantum information processing, particularly from rapid advances in the design of molecular qubits \cite{northwestern}. Metal-organic frameworks (MOFs; also known as coordination polymers) \cite{MOFbook} are an important class of material where there has been spectacular recent progress in designing specific structures and functionalities. MOFs are characterized by a periodic array of metallic ions joined through organic ligands that are chemically bonded to multiple metals, cf. Fig. \ref{fig:WF}.

Much of the recent progress in MOFs has been driven by potential applications such as carbon capture, hydrogen storage, and catalysis \cite{MOFbook}. However, electronic correlations are strong in both organic molecules and transition metals \cite{RRP}. Thus, there has also been significant interest in creating and studying magnetic materials, for example, by synthesizing MOFs containing metal ions with partially filled orbitals \cite{Goddard,Halder}. 
The variety and chemical flexibility of organic ligands that are suitable for MOFs opens up significant possibilities for the rational design of structures. For example, the locations in which ligands chelate metals is well understood and this leads to predictable local structures, particularly for multidentate ligands, and allows for the realization of desired local structural motifs.  
Spin-orbit coupling can be important in  organic molecules \cite{WinterJACS}, organometallic clusters \cite{AmiePRB,JackoPRB17} and heavy metals \cite{Jackeli}, so MOFs also also provide a natural route to realizing anisotropic interactions \cite{JaimePRB16,MerinoPRB17,Oshikawa}.
However, a detailed understanding of how magnetic interactions are mediated by the ligands in MOFs is still lacking. 

\begin{figure}
	\begin{center}
		\includegraphics[width=0.9\columnwidth]{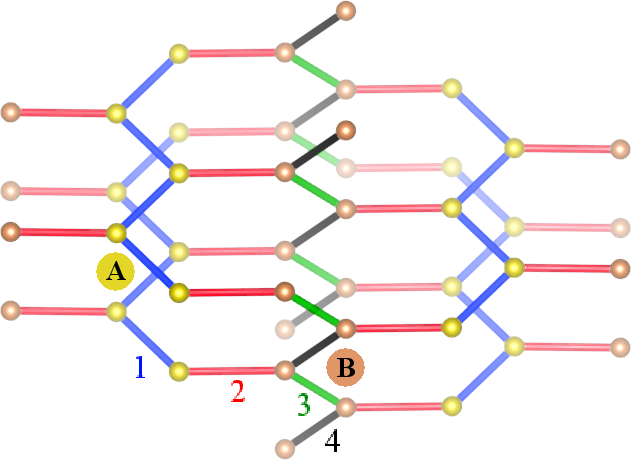} 
		\caption{The lattice of Cu sites (spheres) and C$_2$O$_4$ bridging ligands (lines) in \hhc showing the hyper-honeycomb structure. The crystallographically distinct Cu sites are color coded and labeled $\phi_i=A$ (yellow) or $B$ (copper); the ligands are color coded and labeled $n_{ij}=1-4$. The triethylamine [(C$_2$H$_5$)$_3$NH] counter-ions are not shown.}\label{fig:crystal}
	\end{center}
\end{figure}

In this context, the recent evidence \cite{Pratt18} for a spin liquid groundstate in a MOF, \hhc, is extremely exciting. 
The Cu atoms in \hhc form a hyper-honeycomb or (10,3)-b \cite{Wells} lattice, as sketched in Fig. \ref{fig:crystal}; the copper sites are bridged by oxalate ligands, with interstitial triethylamine molecules. 
No signs of long range order are observed in the heat capacity, magnetic susceptibility, or zero-field muon spin relaxation even at the lowest temperatures studied (60~mK). 
Zhang \textit{et al}.  \cite{Pratt18} argued from an orbital analysis and the inter-site distances that \hhc is described by a Heisenberg model on a quasi-two dimensional `quasi-honeycomb' lattice and that resonating valence bond \cite{AndersonRVB,Savary,RRP} physics is relevant.

In this Letter we systematically construct an effective low-energy model of \hhc. We begin by calculating the electronic structure from density functional theory (DFT). We then construct maximally localised Wannier functions from the Kohn-Sham wavefunctions. We find that these consist predominately  of the Jahn-Teller distorted Cu $e_g$ orbitals. We derive a three-dimensional tight-binging model from the overlaps of the Wannier orbitals and add Kanamori interactions. In the strong coupling limit a Heisenberg model is constructed that consists of a strongly dimerized chain coupled to isotropic Heisenberg chains. On integrating out the strongest interactions (those within the dimers) we are left with a model consisting of very weakly coupled spin chains. We propose that this can explain the quantum spin liquid in \hhc. 

The hyper-honeycomb structure of \hhc is illustrated  Fig. \ref{fig:crystal}. Copper atoms are located at the nodes of the lattice and the bridging oxalate (C$_2$O$_4$) ligands along the bonds. 
The copper atoms are in an approximately octahedral environment; whose symmtery is broken by a Jahn-Teller distortion. The copper atoms are in the Cu$^{2+}$ state, d$^9$ filling. In an octahedral crystal field this would lead to a full t$_{2g}$ shell and a 3/4 full e$_g$ shell. The Jahn-Teller distortion, which is different for the two symmetry distinct types of Cu sites, lifts the degeneracy of the $e_g$ orbitals.

We performed density functional calculations of the electronic structure  in an all-electron full-potential local orbital basis code, FPLO, \cite{koepernik99} using the generalized gradient approximation \cite{perdew96} and scalar relativistic corrections. The density was converged on a $(8 \times 8 \times 8)$ $k$ mesh.

\begin{figure*}
\begin{center}
\includegraphics[width=1.8\columnwidth]{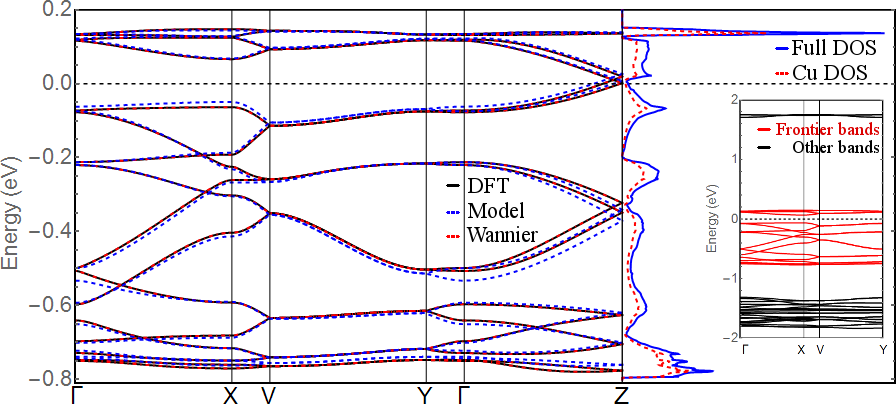} 
\caption{The sixteen frontier bands of \hhc  (left), and the density of states (right). The total density of states (solid blue) is almost totally made up of the contributions of the copper atoms (dashed red). The DFT bands (black) are perfectly represented by the Wannier bands (dashed red) and acurrately described by the nearest neghbour tight-binding model Eq. (\ref{eq:tb}) (blue dashed). The inset shows the band structure in a wider energy window. Observe that the frontier bands are well separated from the other bands.}\label{fig:bs}
\end{center}
\end{figure*}

The computed electronic structure and partial density of states (PDOS) for \hhc is shown in Fig. \ref{fig:bs}. The sixteen bands nearest the Fermi energy (henceforth the frontier bands) arise, to a very large degree, from the two copper 3d-$e_g$ orbitals on each of the eight copper atoms in the unit cell. 
The triethylamine [(C$_2$H$_5$)$_3$NH] counter-ions and oxalate  bridging ligands contribute very little density of states in this energy window.

From the Kohn-Sham orbitals we constructed maximally localized Wannier orbitals from the sixteen frontier bands. 
They are copper centred orbitals with asymmetric extent onto the organic scaffold (Fig. \ref{fig:WF}). By directly calculating the matrix elements of the Hamiltonian between these Wannier orbitals, we produced an \textit{ab initio} nearest-neighbor tight-binding model for  \hhc:
\begin{eqnarray}
\hat{H}_{tb} &=& \sum_{i\sigma} \hat{\bm d}_{i\sigma}^\dagger M_{\phi_i} \hat{\bm d}_{i\sigma} 
+ \sum_{ ij\sigma}\left(\hat{\bm d}_{i\sigma}^\dagger T_{n_{ij}} \hat{\bm d}_{j\sigma} +h.c.\right),
\label{eq:tb}
\end{eqnarray}
where $\hat{\bm d}_{i\sigma}^\dagger=(\hat d_{i1\sigma}^\dagger,\hat d_{i2\sigma}^\dagger)$, $\hat d_{im\sigma}^{(\dagger)}$ annihilates (creates) an electron with spin $\sigma$ in the $m$th Wannier orbital centered on the $i$th Cu atom, $\phi_i\in\{A,B\}$ denotes the sublattice that the $i$th Cu atom belongs to, $n_{ij}$ labels the crystallographically distinct pairs of Cu atoms, and $T_{n_{ij}}\ne0$ only for $n_{ij}=1-4$, the `bonds' marked in Fig. \ref{fig:crystal}. The matrices $M_{\phi_i}$ and $T_{n_{ij}}$ are specified below.

Each copper atom is described by a two-dimensional local Hamiltonian that depends on its crystallographic location. Explicitly the Wannier overlaps yield (in meV)
\begin{equation}
M_A = 
\begin{pmatrix}
-379 & -152\\
-152 & -272
\end{pmatrix};~~
M_B = 
\begin{pmatrix}
-422 & -95\\
-95 & -238
\end{pmatrix}.
\end{equation}
The diagonal elements are the site energies of the two orbitals, and off-diagonal elements are the inter-orbital hopping. 
Trivially, these matrices can be diagonalized by unitary transformations; this would violate the maximum localisation condition and does not change the physics.

Each copper is connected to three other copper atoms via oxylate bridging ligands. Due to the Jahn-Teller distortion of the copper centres, there are four symmetry inequivalent types of oxalate ligand. 
Thus, our tight-binding model contains four distinct  nearest-neighbor hopping matrices, cf. Fig. \ref{fig:crystal}. Each of these is described by a set of four hopping parameters, connecting two orbitals on one site to two on another. They are (in meV)
\begin{eqnarray} 
&T_1 = 
\begin{pmatrix}
 -28  & 91 \\
 121 & 172
\end{pmatrix};&~~ 	
T_2 = 
\begin{pmatrix}
 249  & 37 \\
 -45 & 38
\end{pmatrix};\notag \\	
&T_3 = 
\begin{pmatrix}
 36  & -126 \\
 -126 & 143
\end{pmatrix};&~~	
T_4 = 
\begin{pmatrix}
 35  & 103 \\
 103 & 202
\end{pmatrix}.
\label{eq:T}		
\end{eqnarray}
The largest next-nearest neighbour hopping, neglected in the model [Eq. (\ref{eq:tb})], is $\mathcal{O}(10)$ meV.
Fig. \ref{fig:bs} shows that this nearest-neighbor tight-binding model reproduces the calculated DFT band structure well.

However the DFT predicts that \hhc is a metal (all be it with a small Fermi surface), whereas experimentally it is a magnetic insulator \cite{Pratt18}.
Presumably this indicates that electronic correlations are important. 
The two Cu-$e_g$ orbitals contain three electrons on average, therefore in the insulating state one expects that each pair of $e_g$ orbitals contains three electrons.  We model the electronic interactions on each atom by an $e_g$ symmetry Kanamori interaction \cite{Kanamori,Georges}
\begin{eqnarray}
\hat{V}_K^{e_g} &=& 
U \sum_{im} n_{im\uparrow} n_{im\downarrow} + U' \sum_{i,m \neq m'}n_{im \uparrow} n_{im' \downarrow} \nonumber \\
&&+ (U'-J_H) \sum_{i,m < m',\sigma}n_{im \sigma} n_{im' \sigma} \nonumber \\
&&- J_H \sum_{i,m \neq m'} d_{im\uparrow}^\dagger d_{im\downarrow} d_{im'\downarrow}^\dagger d_{im'\uparrow} \nonumber \\
&&+ J_H \sum_{i,m \neq m'} d_{im\uparrow}^\dagger d_{im\downarrow}^\dagger d_{im'\downarrow} d_{im'\uparrow},
\end{eqnarray}
where $U$ ($U'$) is the intra-(inter-)orbital Coulomb repulsion, and $J_H$ is the Hunds rule coupling. 
For simplicity, we approximate the on-site inter-orbital Coulomb repulsion as $U' = U-2J_H$, which is exact in octahedral symmetry.

\begin{figure}
\begin{center}
\includegraphics[width=0.85\columnwidth]{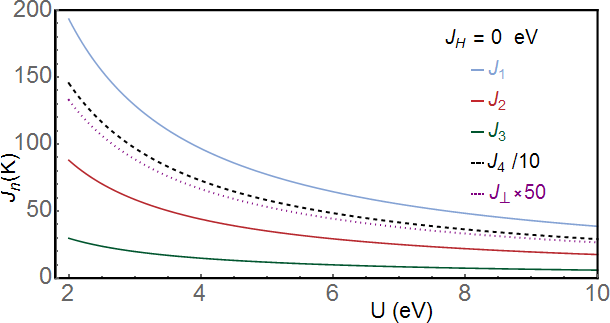} 
\includegraphics[width=0.85\columnwidth]{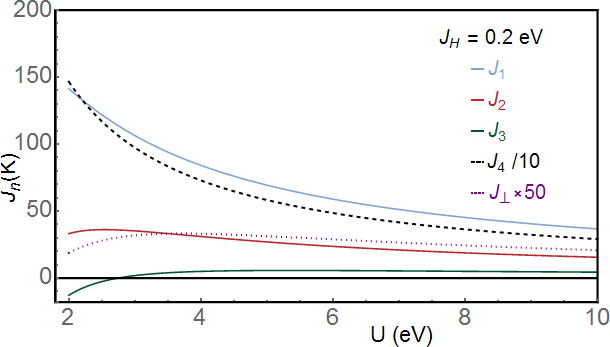} 
\includegraphics[width=0.85\columnwidth]{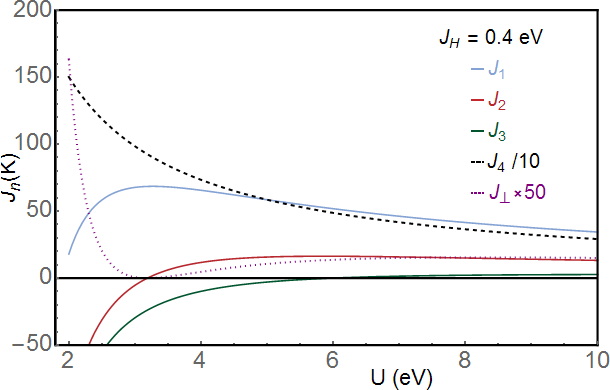} 
\caption{Variation of the effective exchange interactions $J_1$ through $J_4$ and $J_\perp$ with $U$ and $J_{H}$. For reasonable parameters there is a clear hierarchy of energy scales: $J_4\gg J_1 > J_2 \gg J_3 \gg J_\perp$. Note that,  in this figure, $J_4$ is scaled down by a factor of ten and $J_\perp$ is multiplied by fifty.}\label{fig:Js}
\end{center}
\end{figure}

In the Mott insulating phase each pair of $e_g$ orbitals contains three electrons. We perform perturbation theory about this strongly correlated limit to second order in the $\hat{T}$ matrices via a canonical transformation (as described in detail in \cite{PowellPRB17}). 
This results in an effective Heisenberg model 
\begin{equation}
\hat{H}_\text{eff}= J_{n_{ij}} \sum_{ ij}\hat{\bm S}_{i} \cdot \hat{\bm S}_{j},
\label{eq:Heff}
\end{equation}
where $\hat{\bm S}_{i}$ is the spin operator acting on the ground state doublet of an isolated pair of Wannier orbitals.

The calculated effective Heisenberg coupling constants are reported in Fig. \ref{fig:Js}. 
The precise values of $U$ and $J_H$ are not known for \hhc, but as the frontier bands have such strong Cu $e_g$ character we expect values typical of copper in other materials, i.e., $U$ of a few eV and $J_H$ significantly smaller. Even without precise values for $U$ and $J_H$  several important conclusions can be drawn from these results. 
$J_4$ is an order of magnitude greater than the other effective exchange interactions.  $J_3$ is by far the weakest of the four interactions retained in our model for reasonable values of $U$ and $J_H$. Indeed depending on the details of the parameters of the Kanamori interaction $J_3$ may favor either ferromagnetic or antiferromagnetic correlations. For reasonable parameters, our calculated values of the exchange interactions agree well with previous empirical  estimates \cite{Pratt18}.

The orders of magnitude variation of the exchange interactions along different bonds is a direct consequence of the Jahn-Teller distortion of Cu ions and the consequent breaking of the symmetry  of the $e_g$ orbitals.  Hopping sequentially between  three orbitals (one on one site and two on another) can favor either ferromagnetic or antiferromagnetic exchange, depending on the net sign of the hopping integrals \cite{PowellPRL17,PowellPRB17}. 
These interference effects are enhanced by the Hunds rule coupling \cite{MerinoPRB17} and open the possibility of ferromagnetic effective exchange couplings at small $U$.  This also gives a natural explanation of the large difference in the magnitudes of $J_3$ and $J_4$. Note that the off-diagonal terms in $T_3$ are both negative whereas both off-diagonal elements in $T_4$ are positive. Because of the importance of interference effects this has an enormous effect on effective exchange interactions mediated by these hopping matrices. This is straightforwardly checked by calculating the superexchange interactions with the signs of the off-diagonal elements of the $T$ matrices reversed (note that this is not equivalent to any gauge transformation). This decreases $J_4$ by an order of magnitude and increases $J_3$ by an order of magnitude or more (see Fig. S1 \cite{SI}).

The strong spatial anisotropy in the effective exchange interactions leads naturally to a model for the spin liquid behavior observed in \hhc. Observe, Fig. \ref{fig:crystal}, that the crystal can be viewed as consisting of coupled chains of the two crystallographically distinct species of Cu atom (A and B). The strongest and weakest interactions, $J_4$ (black lines in Figs. \ref{fig:crystal} and \ref{fig:Js}) and $J_3$ (green lines in Figs. \ref{fig:crystal} and \ref{fig:Js}) respectively, alternate along chains of B-Cu atoms in the crystal. The exchange interactions along the chains of A-Cu atoms are uniform and of  intermediate magnitude $J_1$ (blue lines in Figs. \ref{fig:crystal} and \ref{fig:Js}). The interchain coupling, $J_2$ (red lines in Figs. \ref{fig:crystal} and \ref{fig:Js}) is significantly smaller than $J_1$. 

\begin{figure}
	\begin{center}
		\includegraphics[width=0.49\columnwidth]{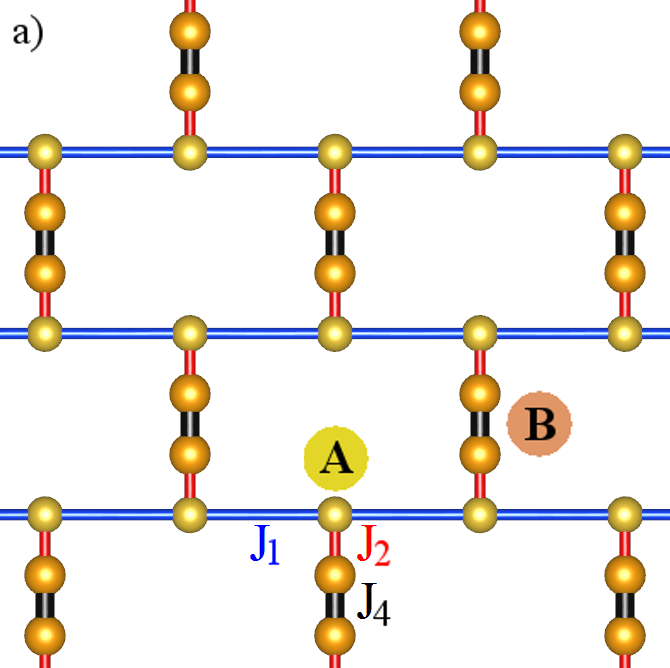}
		\includegraphics[width=0.49\columnwidth]{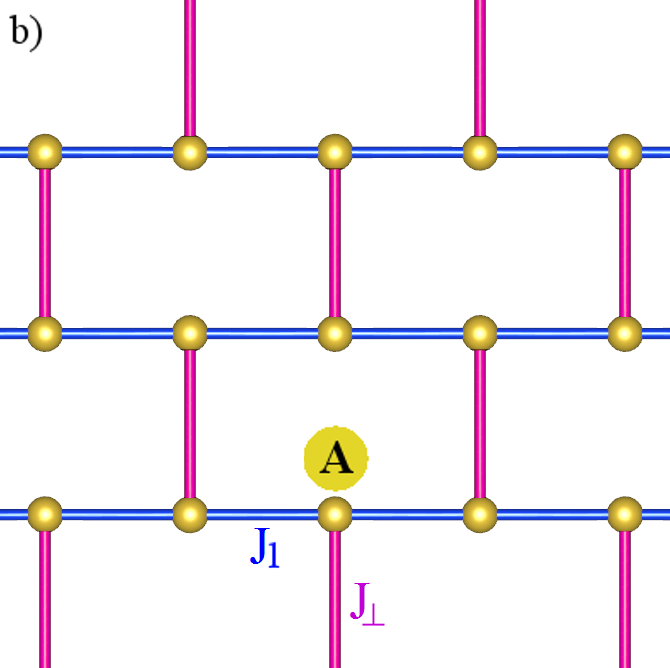} 
		\caption{(a) Sketch of the two-dimensional model that arises for $J_3=0$. Color coding of bonds and Cu atoms as in Fig. \ref{fig:crystal}. (b) The model once the strong ($J_4$) interactions are integrated out. Only the B-sublattice of Cu atoms remains. The same color coding is used with the weak interchain interactions, $J_\perp=J_2^2/2J_4$, represented by the pink lines.}\label{fig:sheet}
	\end{center}
\end{figure}

Given this separation of energy scales it is natural to first consider $J_4\rightarrow\infty$. The B sublattice is dimerized, due to the formation of singlets along the $n_{ij}=4$ bonds. The monogamy of entanglement then guarantees the absence of correlations between spins on the A and B sublattices. Thus, the A-Cu would form isolated Heisenberg chains, with Luttinger liquid ground states \cite{Giamarchi}. This would give a natural qualitative explanation of many of the features observed in experiments on \hhc, such as the linear term in the specific heat \cite{Pratt18,Giamarchi}.

For the realistic problem of large but finite $J_4$ we must consider the role of virtual triplet excitations of the dimers. Given the extreme smallness of $J_3$ we  set $J_3=0$, cf. Fig. \ref{fig:crystal}. This yields a model that is topologically equivalent to a series of uncoupled two-dimensional honeycomb sheets, sketched in Fig. \ref{fig:sheet}a. From this model, one can integrate out the highest energy interaction, $J_4$. Straightforward second order perturbation theory yields a model of chains coupled by a much weaker interaction $J_\perp = J_2^2/2J_4$ (see Fig. \ref{fig:sheet}b), a sub-Kelvin energy scale.

Such models can be treated  using Luttinger liquid theory for the chains and treating interchain interactions via the random phase approximation \cite{Schulz}. As the effective lattice, Fig. \ref{fig:sheet}b, is bipartite, this yields a N\'eel temperature $T_N\propto J_\perp$, consistent with the absence of long range order in \hhc down to 60 mK. Symmetry implies that all further neighbour intrachain interactions on the A sublattice, Fig. \ref{fig:sheet}b, are frustrated, so these interactions, neglected in our model, should not change this result \cite{PowellPRL17,Boquet}. 

The interchain coupling implies that the bosonized Hamiltonian of the chains is a sine-Gordon model, rather than the quadratic Luttinger Hamilontian \cite{Giamarchi}. This, or the physics of the dimerized chains, could be implicated in Zhang \textit{et al},'s finding the the Wilson ratio is less than unity \cite{Pratt18}. However, a detailed calculation of this ratio is beyond the scope of the current Letter.

\begin{acknowledgments}
	We thank Francis Pratt for helpful conversations. This work was supported by the Australian Research Council through grant DP160100060.
\end{acknowledgments}

\section{Supplementary Information}

\begin{figure}
	\begin{center}
		\includegraphics[width=0.85\columnwidth]{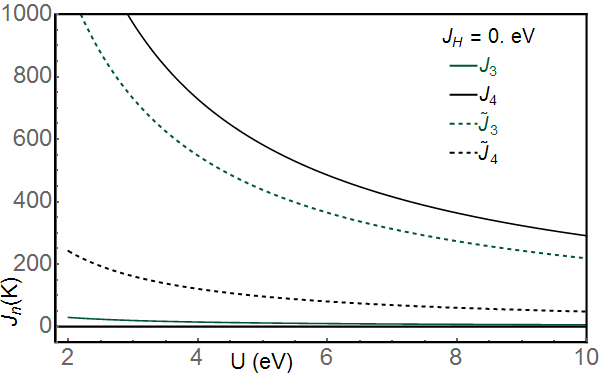} 
		\includegraphics[width=0.85\columnwidth]{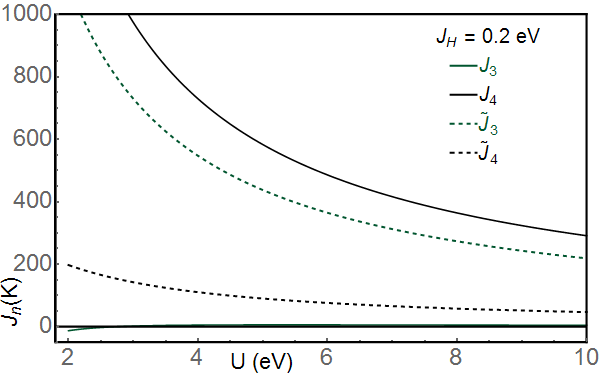} 
		\includegraphics[width=0.85\columnwidth]{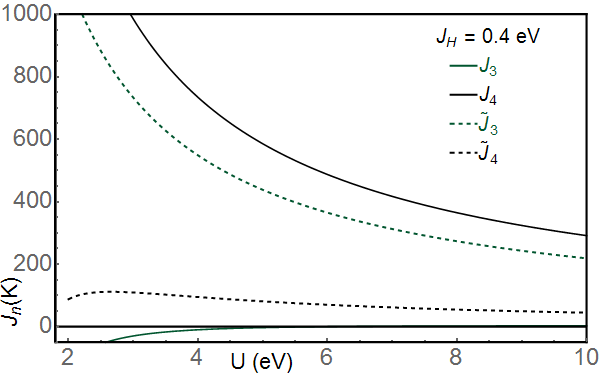} 
		\caption{Effect of the signs of the off-diagonal elements of the $T$ matrices. $\tilde{J}_n$ is the exchange interaction calculated for the same $T$ matrix as $J_n$ except with the signs of the off-diagonal elements reversed [compare Eqs. (\ref{eq:T}) and (\ref{eq:tildeT})]. This changes the values of the exchange interactions by at least an order of magnitude.}\label{fig:sup}
	\end{center}
\end{figure}

We define the matrices 
\begin{eqnarray} 
&\tilde{T}_3 = 
\begin{pmatrix}
36  & 126 \\
126 & 143
\end{pmatrix}&~\text{and}~	
\tilde{T}_4 = 
\begin{pmatrix}
35  & -103 \\
-103 & 202
\end{pmatrix}		
\label{eq:tildeT}
\end{eqnarray}
by reversing the signs of the off-diagonal elements of $T_3$ and $T_4$ respectively. We then calculate the concomitant exchange interactions, $\tilde{J}_3$ and $\tilde{J}_4$ as described in the main text. Fig. \ref{fig:sup} shows that this leads to order of magnitude or larger changes in the values of the exchange integrals; and that $J_3 \ll \tilde{J}_3$ whereas $J_4 \gg \tilde{J}_4$. This confirms that the sign of these elements is the key factor in determining the magnitude of the effect exchange integrals.

\end{document}